\documentclass[superscriptaddress]{revtex4}
\usepackage{graphicx,amsmath}
\usepackage{epsfig}

\begin{document}

\preprint{}

\title{The young center of the Earth}
\author{U.I. Uggerh{\o}j}
\affiliation{Department of Physics and Astronomy, Aarhus University, Denmark}
\author{R.E. Mikkelsen}
\affiliation{Department of Physics and Astronomy, Aarhus University, Denmark}
\author{J. Faye}
\affiliation{Department of Media, Cognition and Communication, University of Copenhagen, Denmark}

\date{\today}
\begin{abstract}
We treat, as an illustrative example of gravitational time dilation in relativity, the observation that the center of the Earth is younger than the surface by an appreciable amount.
Richard Feynman first made this insightful point and presented an estimate of the size of the effect in a talk; a transcription was later published in which the time difference is quoted as 'one or two days'.
However, a back-of-the-envelope calculation shows that the result is in fact a few years.
In this paper we present this estimate alongside a more elaborate analysis yielding a difference of two and a half years.
The aim is to provide a fairly complete solution to the relativity of the 'aging' of an object due to differences in the gravitational potential. This solution - accessible at the undergraduate level - can be used for educational purposes, as an example in the classroom.
Finally, we also briefly discuss why exchanging 'years' for 'days' - which in retrospect is a quite simple, but significant, mistake - has been repeated seemingly uncritically, albeit in a few cases only. The pedagogical value of this discussion is to show students that any number or observation, no matter who brought it forward, must be critically examined.
\end{abstract}

%\maketitle must follow title, authors, abstract, \pacs, and \keywords
\maketitle

\section{Introduction}
The gravitational potential influences the rate at which time passes. 
This means that a hypothetical measurement of the age of a massive object like the Sun or the Earth would yield different results depending on whether performed at the surface or near the center. 
In this connection, clearly, issues such as the initial assembly of cosmic dust to form the protoplanet eventually leading to the Earth is not what is alluded to when considering the age. Rather, the age is understood as e.g.\ the 'aging' of radioactive elements in the Earth, i.e.\ that fewer radioactive decays of a particular specimen have taken place in the Earth center than on its surface.
Furthermore, arguments based on symmetry will convince most skeptics, including those from 'the general public', that there is no gravitational force at the Earth center. Consequently, such an effect cannot be due to the force itself, but may instead be due to the 'accumulated action of gravity' (a layman expression for the gravitational potential energy being the radial integral of the force). 
Thus, there is also a good deal of pedagogical value in this observation. 

In a series of lectures presented at Caltech in 1962-63, Feynman is reported to have shared this fascinating insight with the audience using the formulation "...since the center of the earth should be a day or two younger than the surface!"~\cite{Hatf03}. This thought experiment is just one among a plethora of fascinating observations about the physical world provided by Richard Feynman.
Although this time difference has been quoted in a few papers, either the lecturer or the transcribers had it wrong; it should have been given as 'years' instead of 'days'.

In this paper, we first present a simple back-of-the-envelope calculation which compares to what may have been given in the lecture series. 
We then present a more elaborate analysis which brings along a number of instructive points. 
We believe that this correction only makes the observation of age difference due to gravity even more intriguing. 

We stress that this paper is by no means an attempt at besmearing the reputation of neither Feynman nor any of the authors who trustingly replicated his statement (including one of the authors of the present paper, UIU).
Instead the, admittedly small, mistake is used as a pedagogical point much like the example 'the human failings of genius' that Ohanian has used in his book about Einstein's mistakes \cite{Ohan08}. 
Realising that even geniuses make mistakes may make the scientist more inclined towards critically examining any postulate on his/her own.

\section{The center of the Earth is younger than its surface}

\subsection{Homogeneous Earth}
We initially suppose that the object under consideration is a sphere with radius $R$ and mass $M$, homogeneously distributed.
Its gravitational potential as a function of distance $r$ to its center is then given by
\begin{eqnarray}
\Phi&=&-G\cfrac{M}{r}~~~~~~~~~~~~~~~~~~~r\geq R\\
\Phi&=&-G\cfrac{M(3R^2-r^2)}{2R^3}~~~~~~r\leq R
\label{earth_time1}
\end{eqnarray}
such that the potential on its surface is $\Phi(R)=-GM/R$ and the potential in its center is $\Phi(0)=-3GM/2R$.
The difference between the gravitational potential at the center and at the surface is then
\begin{equation}
\Delta\Phi=\Phi(R)-\Phi(0)=\cfrac{1}{2}G\cfrac{M}{R}.
\label{earth_time4}
\end{equation}
A difference in gravitational potential implies a time dilation at the point with the lower potential.
This is given by the standard 'gravitational redshift'
\begin{equation}
\omega=\omega_0(1-\cfrac{\Delta\Phi}{c^2}),
\label{work_height6}
\end{equation}
which here relates the (angular) frequencies at the center, $\omega$, and at the surface $\omega_0$. Being the inverse of the period, the frequency is indirectly a measure of how quickly time passes. It is customary to use the symbol $\omega$ in this connection, and we emphasise that this variable has nothing to do with the Earth rotation.

We combine equation \eqref{work_height6} with the result for $\Delta\Phi$ in equation \eqref{earth_time4} and use that $\Delta\omega=\omega-\omega_0$,
\begin{equation}
\Delta\omega=-\cfrac{1}{2}G\omega_0\cfrac{M}{Rc^2}
\label{earth_time5}
\end{equation}
We note that this treatment is based on equation \eqref{work_height6} which "...refers only to identically constructed clocks located at different distances from the center of mass of a gravitating body along the lines of force. All that is required is that the clocks obey the weak equivalence principle [...] and the special theory of relativity." \cite{Nobi13}. 
See Ref.~\cite{RSChristensen} for a recent, instructive example that can be easily performed in the undergraduate laboratory to display one aspect of the equivalence principle.

For the case of the Earth, upon rewriting and setting the surface acceleration $GM_e/R_e^2=g=9.82$ m/s$^2$, with $R_e$ being the Earth radius, equation \eqref{earth_time5} becomes
\begin{equation}
\cfrac{\Delta\omega}{\omega_0}=-\cfrac{1}{2}\cfrac{R_eg}{c^2},
\label{earth_time6}
\end{equation}
such that the Earth mass $M_e$ and the gravitational constant $G$ are not needed explicitly.

For the sake of a back-of-the-envelope calculation we may exploit that $c/g\simeq1$ year (within 3\%, although there is no direct connection between the motion of the Earth around the Sun and $c$). The Earth age is $T_e=4.54\cdot10^9$ years and its average radius is $R_e=$ 6371 km so that $R_e/2c$ is approximately 10 ms. A year is approximately $\pi\cdot10^7$ s. Clearly, the use here of $\pi$ is a mnemonic device, not an expression of precision, although it is precise to about half a percent (one could use $3$ instead of $\pi$ which, however, is imprecise to 5 percent). Thus the difference between the age of the Earth surface and its center becomes approximately $4.5\cdot10^9\cdot10^{-2}/\pi\cdot10^7\simeq1.4$ years, with the center being youngest.
This is the \emph{type} of 'back-of-the-envelope' calculation that one could imagine that Feynman had in mind when he expressed his "...since the center of the earth should be a day [which thus should have read 'year'] or two younger than the surface!" \cite{Hatf03}. Where the mistake actually entered in the lecture and transcription process is unlikely to ever be ascertained, and its exact origin is not important for the following discussion.

With tabulated values for $M_e,~G,~R_e,~c$ and $T_e$ a more precise number for the homogeneous Earth is obtained:
\begin{equation}
\Delta T_{eh}=T_e\cfrac{1}{2}G\cfrac{M_e}{R_ec^2}=1.58~\mathrm{years},
\label{earth_time7}
\end{equation}
with the center being youngest.

\subsection{Realistic Earth}
Rather than assuming a homogeneous Earth, we now turn to a more realistic density distribution. 
This yields a significantly different result and reveals some insights to the origin of the time difference. 
A rather precise description, but not the only one available, of the Earth density profile is tabulated in the so-called 'Preliminary Reference Earth Model' (PREM)~\cite{Dzie81}. 
Very recently, the PREM has been applied to give a detailed description of the Earth 'gravity tunnel' problem \cite{Klot15}.

\begin{figure}[hbt]
\begin{center}
\includegraphics[width=0.5\textwidth]{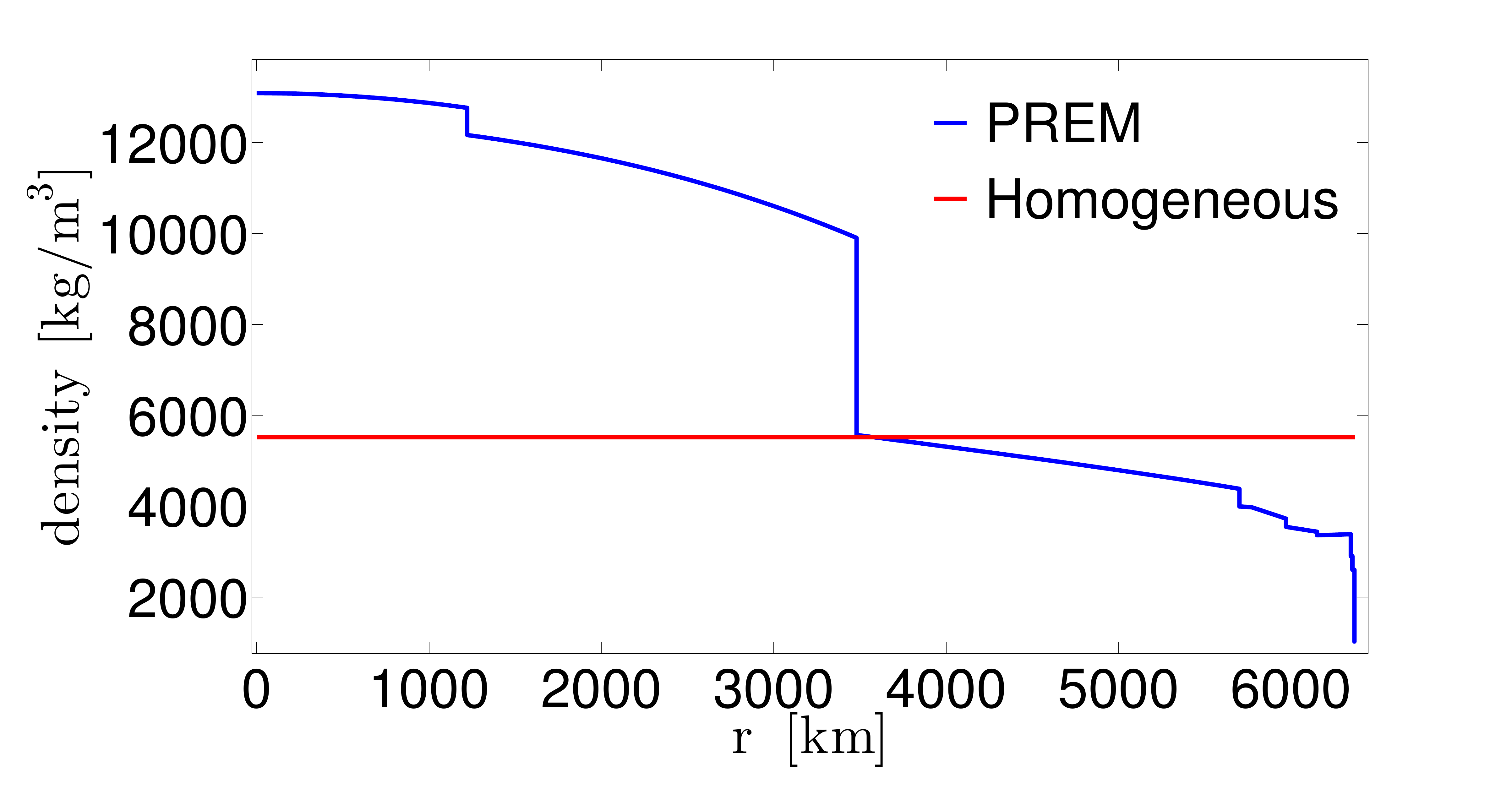}
\caption{The density of the Earth as a function of distance to the Earth center for two different models. The blue line shows the PREM of the Earth density and the red curve is the constant density in the approximation of a homogeneous Earth with mass $M_e$.}
\label{fig:Density}
\end{center}
\end{figure}

We shall consider a spherically symmetric Earth with a density only dependent on radius, $\rho (r)$, as given by the PREM, see Figure \ref{fig:Density}.
The gravitational potential caused by this sphere is then given by
\begin{align}
\Phi(r) = -\int_{\infty}^{r}\vec{f}_{\textrm{grav}}\cdot \vec{dr'},
\end{align}
where $\vec f_{\textrm{grav}}=\vec F_{\textrm{grav}}/m$ is the mass specific force, or acceleration $\vec a_{\textrm{grav}}$, due to gravity, with $\vec F_{\textrm{grav}}$ being the gravitational force (which is why $\Phi$ is the gravitational potential and not the gravitational potential \emph{energy}). The gravitational potential energy is equal to the work done by taking a test particle of mass $m$ from infinity to a distance $r$ away from the center of the Earth.

\begin{figure}[hbt]
\begin{center}
\includegraphics[width=0.5\textwidth]{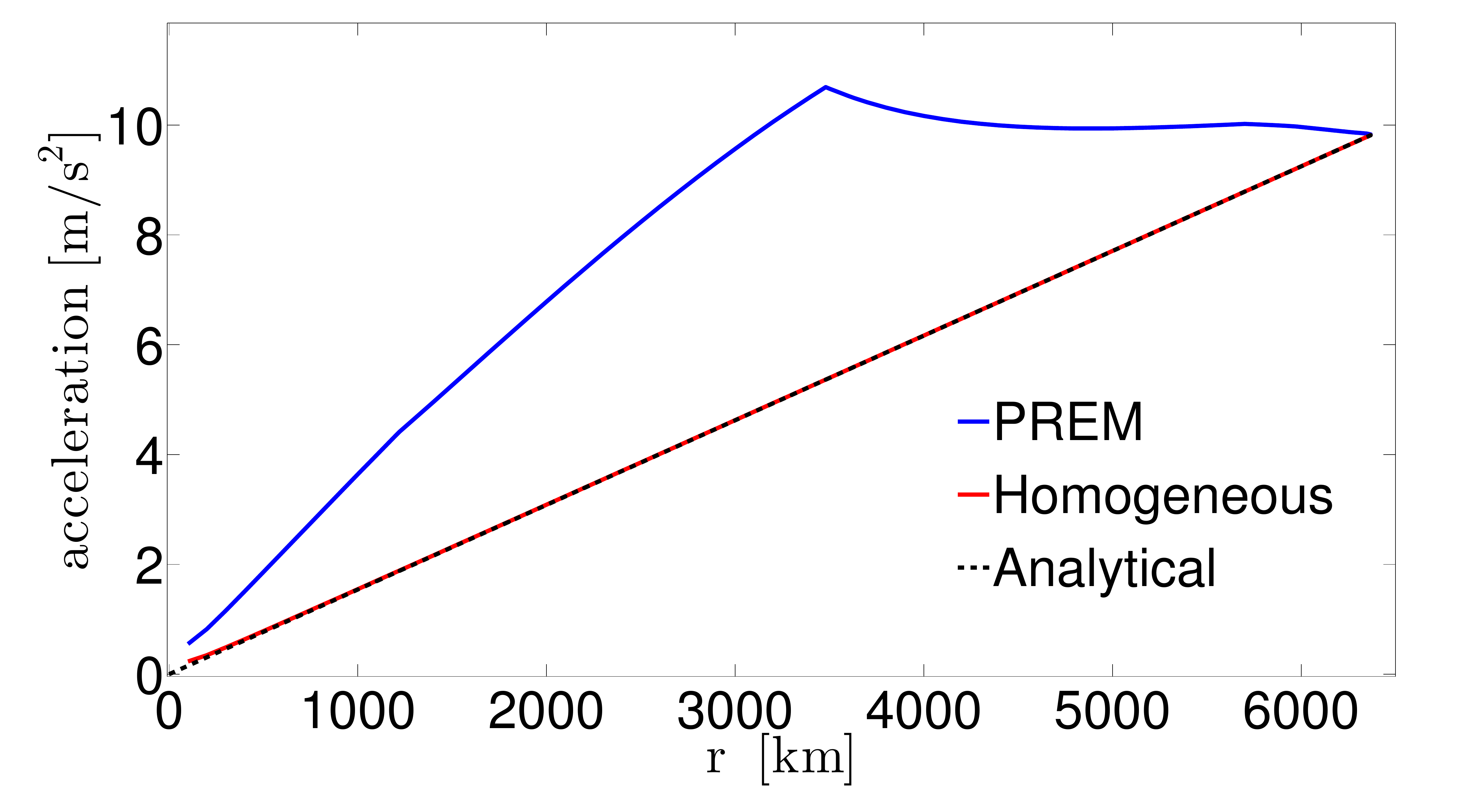}
\caption{The size of the gravitational acceleration as a function of distance to the Earth center. It reaches the familiar value of $9.82$ m/s$^2$ at the surface. The analytical curve is given by the simple scaling $g\cdot r/R_e$ by assuming a homogeneous mass distribution.}
\label{fig:Acceleration}
\end{center}
\end{figure}

We split the expression in two parts:
\begin{align}
\Phi(r) =& -\int_{R}^{r}\vec{f}_{\textrm{grav}}\cdot \vec{dr'} - \int_{\infty}^{R}\vec{f}_{\textrm{grav}}\cdot \vec{dr'},
\end{align}
with the first term being the work per unit mass done inside the object - in this case the Earth - and the last term the work per unit mass done moving the test particle from infinity to the Earth surface.
The gravitational acceleration at a distance, $r$, outside a sphere of mass $M$ is $\vec a_{\textrm{grav}}=\vec f_{\textrm{grav}}=-\hat{r}GM/r^{2}$, the sign showing that it is directed towards the center.
Inside the sphere, when $r < R$, only the mass closer than $r$ to the center matters. We denote this by
\begin{align}
M(r) = \int_0^r 4\pi r'^2 \rho(r')dr'.
\label{M_r}
\end{align}
Now we can write the sum specifically for $r < R$ as
\begin{align}
\Phi(r) =& \int_{R}^{r} G\frac{M(r')}{r'^2} dr' -G\cfrac{M}{R}
\label{Phi_r}
\end{align}
where the last term is the potential at the surface of the object.
The integrand in the first term is the gravitational acceleration as a function of $r$.
When evaluated at the surface, $r=R_e$, the result is the normal gravitational acceleration, $g$.
This can be seen on Figure \ref{fig:Acceleration} where the acceleration felt at different distances to the Earth center is shown. Due to the mass distribution 'kink' seen in Figure \ref{fig:Density} at a radius of about 3500 km, the acceleration becomes almost constantly equal to its surface value from this radius, outwards. 

Using the PREM density distribution $\rho(r')$ in eq.\ \eqref{M_r} as an input to eq.\ \eqref{Phi_r}, the more elaborate result for the age difference of the Earth center and the surface is
\begin{equation}
\Delta T_e=2.49~\mathrm{years},
\label{earth_time7b}
\end{equation}
with the center being youngest.

\begin{figure}[hbt]
\begin{center}
\includegraphics[width=0.5\textwidth]{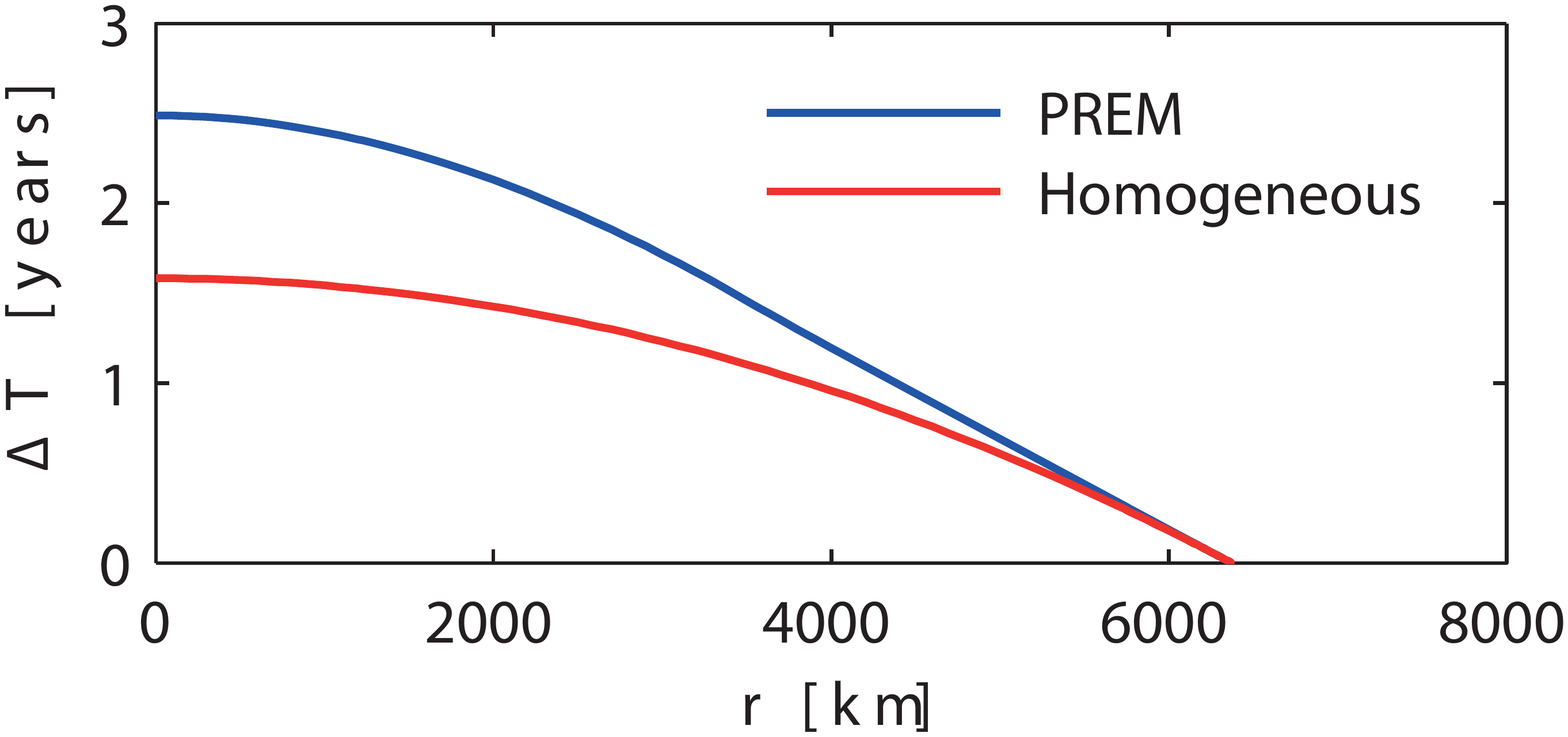}
\caption{The figure shows the time difference between a point inside the Earth and the surface (center) of the Earth. The blue curve is calculated with PREM and the red curve using a homogeneous mass distribution.}
\label{fig:TvariationE}
\end{center}
\end{figure}

As a, perhaps, intriguing side-effect, we show the time difference as a function of radius, see Figure \ref{fig:TvariationE}.
%From the lower part of the figure one observes that, for example, the age (according to this model, and in the above mentioned sense) at the bottom of the Mariana Trench at 11 km depth is found to be around 50 hours less than at the surface.

As expected, the two theories predict similar time differences near the surface of the Earth. Closer to the center, the PREM yields a larger result than the homogeneous distribution.
This is because $M(r)_{\mathrm{PREM}}>M(r)_{\mathrm{Homog.}}$ for small $r$. In fact, assuming for simplicity that the object of radius $R$ consists of a region of high density for $0\leq r\leq r_0$ and zero density for $r_0< r\leq R$, respecting that the total mass equals $M$, and with $r_0=k_0R$, $0<k_0\leq1$, the potential difference between center and surface becomes
\begin{equation}
\Delta\Phi=(\cfrac{3}{k_0}-2)\Delta\Phi_h,
\label{DPhi_hom}
\end{equation}
where $\Delta\Phi_h$ is that of the homogeneous distribution. Thus, the factor $3/k_0-2$ yields the increase in time difference compared to the homogeneous model. So for the Earth, where this approximation is rather crude, we may set $r_0$ to be 3480 km as seen from the PREM curve in Figure \ref{fig:Density}, i.e.\ $k_0\simeq3480~\mathrm{km}/6371~\mathrm{km}\simeq0.55$ such that $\Delta\Phi\simeq3.5\Delta\Phi_h$ somewhat above the factor 1.7 obtained by the numerical method, as expected from the crudeness of this approximation.

We end this section by showing that the time dilation due to the rotational speed of the surface of the Earth makes a negligible contribution. The surface speed is given from the period of rotation as $v_s=R_e 2\pi/T_s\simeq464$ m$/$s where $T_s=86.164,099$ s is the stellar day (the Earth rotation period with respect to the 'fixed' stars). Since the time dilation in special relativity is given from the Lorentz factor $\gamma=1/\sqrt{1-v^2/c^2}\simeq1+v^2/2c^2$ as $\Delta T=Tv^2/2c^2$ we get with $T_e$ that $\Delta T_s\simeq5\cdot10^{-3}$ years, which can be neglected in the present discussion.

\section{The case of the Sun}
Clearly, the calculations performed in connection with the Earth can be performed for essentially any other cosmic object with known mass and radius, at least in the limit of a homogeneous mass distribution. However, we limit the additional cases to our cosmic neighbourhood, i.e.\ to that of the Sun, in order to demonstrate the applicability of eq.\ \eqref{DPhi_hom}. For the Sun, in analogy with the PREM which is based on seismic data, we choose the so-called 'Model S' for its density distribution, a model in good agreement with helioseismic data \cite{Chri96}.
\begin{figure}[hbt]
\begin{center}
\includegraphics[width=0.5\textwidth]{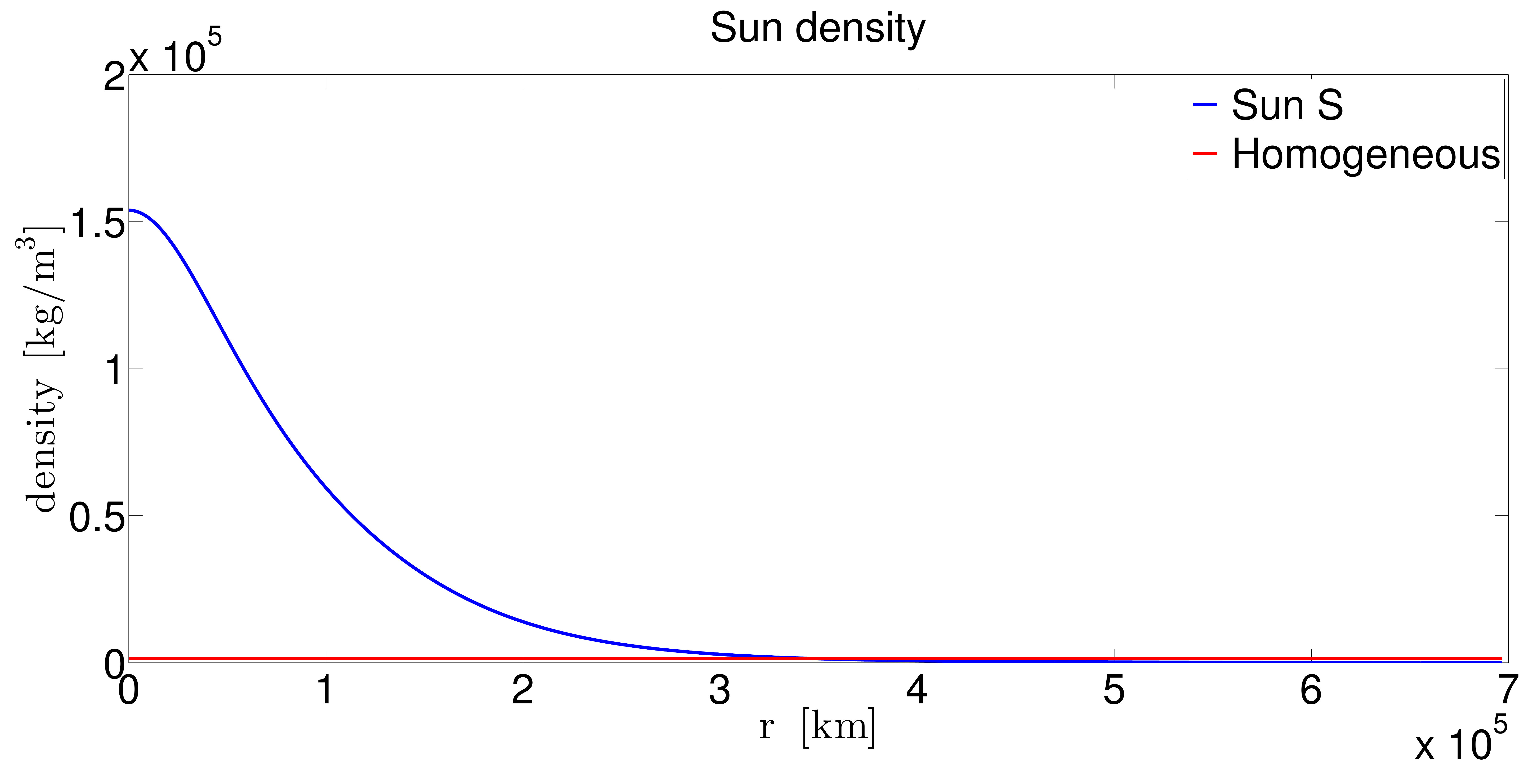}
\caption{The density of the Sun as a function of distance to the Sun center for two different models. The blue line shows the 'Model S' of the Sun density, obtained from helioseismic data \cite{Chri96}, and the red curve is the uniform density Sun.}
\label{fig:DensitySun}
\end{center}
\end{figure}

\begin{figure}[hbt]
\begin{center}
\includegraphics[width=0.5\textwidth]{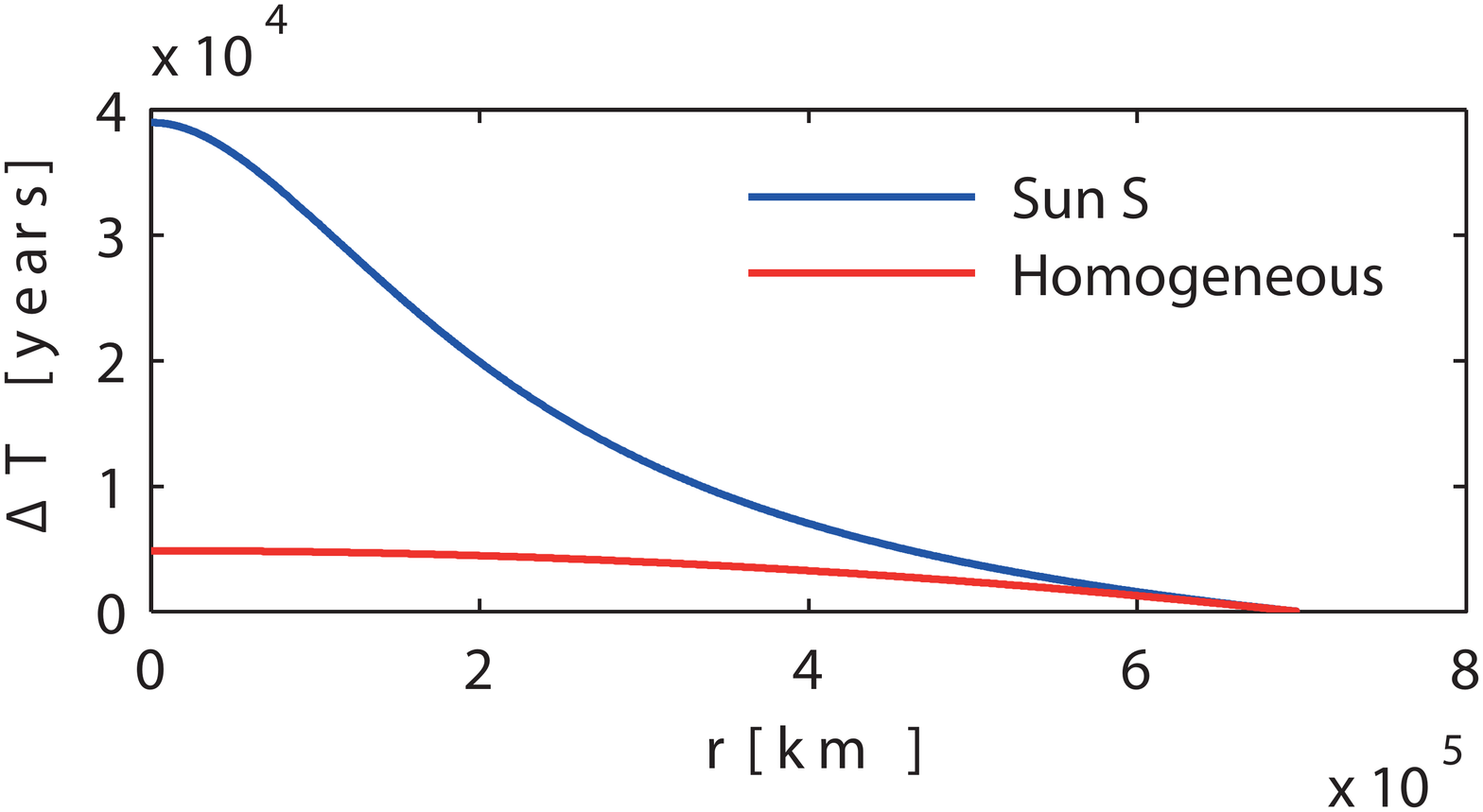}
\caption{The figure shows the time difference between a point inside the Sun and the  surface (center) of the Sun. The blue curve is calculated with the 'Model S' density model for the Sun and the red curve using a homogeneous mass distribution.}
\label{fig:TvariationS}
\end{center}
\end{figure}

In the homogeneous case, the age difference between the Sun center and surface, which can be rewritten as $\Delta T=T_s v_{\mathrm{esc}}^2/4c^2$ with $v_{\mathrm{esc}}=\sqrt{2GM/R}$ being the surface escape velocity, is
\begin{equation}
\Delta T_{sh}=4.8\cdot10^3~\mathrm{years},
\label{sun_time1}
\end{equation}
whereas with the 'Model S' solar model it becomes
\begin{equation}
\Delta T_{s}=3.9\cdot10^4~\mathrm{years},
\label{sun_time2}
\end{equation}
see Figure \ref{fig:TvariationS}.

The factor of $8.0$ difference between these two numbers is substantially larger than that between the same two numbers for the Earth. 
This is a result of the Earth being relatively homogeneous while for the Sun, a significantly larger part of its mass is located close to its center.
Using eq.\ \eqref{DPhi_hom} and approximating $k_0\simeq2.5/7$ from the density distribution, the Model S curve in Figure \ref{fig:DensitySun}, we get a factor $8.4$, a much better approximation than for the case of the Earth.

\section{Discussion and conclusion}

As a final discussion we address the question: why did famous, respectable and clever physicists publish Feynman's claim (although not verbatim, actually) that "[Feynman] concluded that the center of the Earth should be 'a day or two younger than its surface'" \cite{Okun99}, or "[Feynman] concludes that the center of the Earth should be by a day or two younger than its surface" \cite{Okun99b} and reversely "'atoms at the surface of the earth are a couple of days older than at its center'", in the latter case even with the comment "this was confirmed by airplane experiments in 1970s" \cite{Okun02}?
And why did other, equally talented physicists not correct \emph{that} particular mistake in the foreword to the transcribed lectures, in spite of quite extensive discussions, spanning 24 pages, of among other things a few misconceptions etc. \cite{Pres95}? Not to mention the transcribers - postdocs with Feynman - who, along the way, probably have corrected a few mistakes here and there? Or the editor, who also provided introductory notes on quantum gravity \cite{Hatf03b}? Why did one of us (UIU), repeat the same mistake in a science book for the layman \cite{Ugge14}?

This, of course, was not because any of these physicists were unable to check the original claim, or found it particularly laborious to do so. Instead, it seems likely that they knew that the qualitative effect had to be there, and simply trusted that Feynman and his transcribers had got the number right. This is here considered an example of 'proof by ethos' \cite{Faye14}.

%It should be noted, however, that this case of proof by ethos is not the only example. Another well-documented case, which went unnoticed for a long time, is from the formulation of the laws of relativistic thermodynamics. Not so long after Einstein came forward with the special theory of relativity, he and Max Planck applied the theory to the areas of thermodynamics. But even though these great physicists both failed to provide the correct formulas, their results were repeated over the years in many text books of relativity until Heinrich Ott and then in 1965 H. Arzeli\'{e}s independently discovered that the formulas were incorrect \cite{Moll72}.

The term 'proof by ethos' refers to cases where a scientist's status in the community is so high that everybody else takes this person's calculations or results for granted. In other words, nobody questions the validity of that scientist's claim because of the particular ethos that is associated with that person. The result is accepted merely by trust.
Indeed, the proof by ethos is not really a proof as it does not follow logically from a set of premises. But it is a proof in the sense that it is persuasive, and tells us something about how scientists work in practice when they accept a calculation or an experimental result. Scientists must to a large extent rely on the validation of other fellow's work, and it happens to be a psychological default condition among many (scientists), that if a famous peer has publicly announced a result, it is accepted at face value. This seems also to be the situation in the case of the flawed estimate of the relativistic age of the Earth's core. 

In science, one route to becoming famous is being right on some important topics.
However, just because someone has become famous, this person is evidently not necessarily right on all matters.
Feynman himself would most likely have agreed with this and he would probably not have fallen for his own miscalculation: 
For a long time, his own theory of beta decay was at odds with the then prevalent, but false, understanding of existing experimental results. 
Upon finally realizing and correcting this community-wide misunderstanding Feynman wrote: "Since then I never pay any attention to "experts". I calculate everything myself."~\cite{Surely}.
And when faced with a mistake of his own, he put it even more bluntly: "What it says in the book [I have written] is absolutely wrong!" \cite{Hey99}.\\

In spite of the small numerical mistake, Feynman's observation that the center of the Earth is younger than its surface is a fascinating demonstration of time dilation in relativity, and as such a very illustrative example for use in the classroom.

\end{document}